\documentclass[twocolumn]{aastex631}
\usepackage{amsmath}
\usepackage{amssymb}
\usepackage{graphicx}

\shorttitle{A Multi-Wavelength View of SPT-CL J0607-4448}
\shortauthors{Masterson et al.}

\begin{document}

\title{Evidence for AGN-Regulated Cooling in Clusters at $z \sim 1.4$: A Multi-Wavelength View of SPT-CL~J0607-4448}

\author[0000-0003-4127-0739]{Megan Masterson}
\affiliation{MIT Kavli Institute for Astrophysics and Space Research,
Massachusetts Institute of Technology, 
Cambridge, MA 02139, USA}
	
\author[0000-0001-5226-8349]{Michael McDonald}
\affiliation{MIT Kavli Institute for Astrophysics and Space Research,
Massachusetts Institute of Technology, 
Cambridge, MA 02139, USA}

\author[0000-0002-6443-3396]{Behzad Ansarinejad}
\affiliation{School of Physics, 
University of Melbourne, 
Parkville, VIC 3010, Australia}

\author[0000-0003-1074-4807]{Matthew Bayliss}
\affiliation{Department of Physics, 
University of Cincinnati, 
Cincinnati, OH 45221, USA}

\author[0000-0002-5108-6823]{Bradford A. Benson}
\affiliation{Fermi National Accelerator Laboratory, 
Batavia, IL 60510-0500, USA}
\affiliation{Department of Astronomy and Astrophysics, 
University of Chicago, 
Chicago, IL 60637, USA}
\affiliation{Kavli Institute for Cosmological Physics, 
University of Chicago, 
Chicago, IL 60637, USA}

\author[0000-0001-7665-5079]{Lindsey E. Bleem}
\affiliation{High Energy Physics Division, 
Argonne National Laboratory, 
Argonne, IL 60439, USA}
\affiliation{Kavli Institute for Cosmological Physics, 
University of Chicago, 
Chicago, IL 60637, USA}

\author[0000-0002-2238-2105]{Michael S. Calzadilla}
\affiliation{MIT Kavli Institute for Astrophysics and Space Research,
Massachusetts Institute of Technology, 
Cambridge, MA 02139, USA}

\author[0000-0002-3398-6916]{Alastair C. Edge}
\affiliation{Department of Physics,
Durham University,
Durham DH1 3LE, UK}

\author[0000-0003-4175-571X]{Benjamin Floyd}
\affiliation{Department of Physics and Astronomy, 
University of Missouri–Kansas City, 
5110 Rockhill Road, Kansas City, MO 64110, USA}

\author[0000-0001-6505-0293]{Keunho J. Kim}
\affiliation{Department of Physics, 
University of Cincinnati, 
Cincinnati, OH 45221, USA}

\author[0000-0002-3475-7648]{Gourav Khullar}
\affiliation{Department of Astronomy and Astrophysics, University of
Chicago, 5640 South Ellis Avenue, Chicago, IL 60637}
\affiliation{Kavli Institute for Cosmological Physics, University of
Chicago, 5640 South Ellis Avenue, Chicago, IL 60637}
\affiliation{MIT Kavli Institute for Astrophysics and Space Research, Massachusetts Institute of Technology, Cambridge, MA 02139, USA}

\author[0000-0003-3521-3631]{Taweewat Somboonpanyakul}
\affiliation{Kavli Institute for Particle Astrophysics \& Cosmology, 
P.O. Box 2450, Stanford University, 
Stanford, CA 94305, USA}

\correspondingauthor{Megan Masterson}
\email{mmasters@mit.edu}

\begin{abstract}
    We present a multi-wavelength analysis of the galaxy cluster SPT-CL~J0607-4448 (SPT0607), which is one of the most distant clusters discovered by the South Pole Telescope (SPT) at $z=1.4010\pm0.0028$. The high-redshift cluster shows clear signs of being relaxed with well-regulated feedback from the active galactic nucleus (AGN) in the brightest cluster galaxy (BCG). Using \textit{Chandra} X-ray data, we construct thermodynamic profiles and determine the properties of the intracluster medium. The cool core nature of the cluster is supported by a centrally-peaked density profile and low central entropy ($K_0=18_{-9}^{+11}$\,keV\,cm$^2$), which we estimate assuming an isothermal temperature profile due to the limited spectral information given the distance to the cluster. Using the density profile and gas cooling time inferred from the X-ray data, we find a mass cooling rate of $\dot{M}_\mathrm{cool}=100_{-60}^{+90}~M_\odot$~yr$^{-1}$. From optical spectroscopy and photometry around the [\textsc{O~ii}] emission line, we estimate that the BCG star formation rate is SFR$_\mathrm{[O\,\textsc{II}]}=1.7_{-0.6}^{+1.0}~M_\odot$~yr$^{-1}$, roughly two orders of magnitude lower than the predicted mass cooling rate. In addition, using ATCA  radio data at 2.1~GHz, we measure a radio jet power of $P_\mathrm{cav}=3.2_{-1.3}^{+2.1}\times10^{44}$~erg~s$^{-1}$, which is consistent with the X-ray cooling luminosity ($L_\mathrm{cool}=1.9_{-0.5}^{+0.2}\times10^{44}$~erg~s$^{-1}$ within $r_\mathrm{cool}=43$~kpc). These findings suggest that SPT0607 is a relaxed, cool core cluster with AGN-regulated cooling at an epoch shortly after cluster formation, implying that the balance between cooling and feedback can be reached quickly. We discuss implications for these findings on the evolution of AGN feedback in galaxy clusters.
\end{abstract}



\section{Introduction} \label{sec:intro}

A galaxy cluster contains tens to hundreds of member galaxies (with some reaching over a thousand members) surrounded by hot, ionized plasma called the intracluster medium (ICM), all embedded in a massive dark matter halo that constitutes the majority of the cluster mass. The ICM is the dominant baryonic component of clusters, and it is visible at X-ray wavelengths via bremsstrahlung radiation caused by the motion of charged particles. We often classify galaxy clusters into two main groups---cool core clusters, in which the central temperature drops and the density increases, and non-cool core clusters, which have cores that are roughly isothermal. In cool core clusters, the ICM has short radiative cooling times and should produce massive cooling flows of $\dot{M} \sim 100-1000$ $M_\odot$ yr$^{-1}$, in which cold gas condenses out of the hot plasma \citep[see][for a review]{Fabian1994}. However, such cooling flows are not observed in most systems, with typical star formation rates on the order of $\sim1\%$ the expected cooling rate \citep[e.g.][]{ODea2008,McDonald2018} and a lack of cool gas as probed with high resolution X-ray spectroscopy \citep[e.g.][]{Peterson2003,Bregman2006,Peterson2006}. 

One of the dominant mechanisms that is thought to prevent the rapid cooling of the ICM is mechanical feedback from an active galactic nucleus (AGN) in the brightest cluster galaxy \citep[BCG; e.g.][]{McNamara2007,McNamara2012,Fabian2012}. In this paradigm, the radio-loud AGN is accreting well below the Eddington limit and launches powerful jets that inject energy into the ICM by inflating bubbles and thus creating X-ray cavities. Observationally, the inflation of these bubbles has been shown to have enough energy to balance the cooling flow in many systems \citep[e.g.][]{Birzan2004, Dunn2006, Rafferty2006, Hlavacek-Larrondo2012, Hlavacek-Larrondo2015}. Although AGN feedback is now generally accepted as one of the main heating mechanisms balancing cooling in clusters of galaxies, there are still many open questions, including how the properties of the ICM and the impact of AGN feedback have evolved over cosmic time.

The study of high-redshift galaxy clusters and cluster evolution has been greatly aided by recent advances in the millimeter/sub-millimeter regime, whereby the thermal Sunyaev-Zel'dovich (SZ) effect can be used to detect galaxy clusters using their imprint on the cosmic microwave background \citep{Sunyaev1972}. Millimeter observatories like the \textit{Planck} satellite \citep{PlanckCollaboration2016}, the Atacama Cosmology Telescope \citep[ACT;][]{Hilton2018,Hilton2021}, and the South Pole Telescope \citep[SPT;][]{Carlstrom2011, Bleem2015, Bleem2020,Huang2020} have greatly increased the number of detected galaxy clusters at $z > 1$. The SZ selection method is mass-limited, nearly redshift-independent \citep[e.g.][]{Bleem2015}, and independent of the dynamical state of the cluster \citep[e.g.][]{Nurgaliev2017}, allowing for a selection function well-suited for cluster evolution studies. In addition, SZ detection avoids significant bias toward strong cool core systems \citep[e.g.][]{Lin2015}, which plagues X-ray detection mechanisms \citep[e.g.][]{Eckert2011}, and avoids any bias due to cluster galaxy properties that are present in optical and infrared detection methods. 

Uniform X-ray follow-up of SZ-selected clusters has revealed similarity among ICM thermodynamic properties and the impact of AGN feedback on the ICM from $z \sim 0$ up to $z \sim 1.7$ \citep[e.g.][]{McDonald2013,Hlavacek-Larrondo2015,McDonald2017,Ruppin2021,Ghirardini2021}. In particular, the density profiles of clusters are consistent with self-similar evolution in the outskirts and with no redshift evolution in the cores \citep{McDonald2017,Ruppin2021}, indicating consistent non-gravitational processes at play in cluster cores responsible for the deviation from self-similarity. Likewise, \cite{Hlavacek-Larrondo2015} found that the power from AGN feedback in cool core clusters has been roughly constant up to $z \sim 1$. Probing the ICM in the most distant clusters will be a primary focus of next generation X-ray missions like \textit{Athena} \citep{Barret2020}. For now, focusing on multi-wavelength observations of the most distant clusters allows us to place constraints on the nature of AGN feedback and ICM properties at $z > 1$.

SPT-CL~J0607-4448 (hereafter SPT0607) is one of the most distant SPT clusters discovered to date \citep{Bleem2015}, with a redshift of $z = 1.4010 \pm 0.0028$ as measured by spectroscopic follow-up of cluster members \citep{Khullar2019}. As such, it has been extensively followed up with various observatories and has been studied in the X-ray as part of the SPT-SZ high-$z$ sample \citep{McDonald2017, Ghirardini2021}. In the optical band, SPT0607 seems to contain two main groups of galaxies, one at $z = 1.401$ and one closer to $z \sim 1.48$. However, the red sequence, dynamics of the cluster members, and spectroscopy of the BCG favors the lower redshift solution \citep{Khullar2019, Strazzullo2019}. Finally, the galactic properties of cluster members were investigated in \cite{Strazzullo2019}, where they found an overdensity of red galaxies in the cluster, although this overdensity was less prominent than other clusters in their sample (with $1.4 \lesssim z \lesssim 1.7$) despite SPT0607 having the most massive BCG. Our analysis of SPT0607 brings together multi-wavelength observations to put together the full picture of this relaxed, cool core cluster with well-regulated cooling and feedback at such a high redshift.

This work is organized as follows. In Section \ref{sec:obs}, we outline the multi-wavelength data analyzed in this work. We present our results in Section \ref{sec:results} and discuss the implications of these findings on our understanding of cluster evolution and the AGN feedback process at high redshift in Section \ref{sec:discussion}. Finally, we summarize our findings in Section \ref{sec:summary}. Throughout this work, we utilize a $\Lambda$CDM cosmology with $H_0 = 70$ km s$^{-1}$ Mpc$^{-1}$, $\Omega_M = 0.3$, and $\Omega_\Lambda = 0.7$. All quoted uncertainties correspond to 68\% ($1\sigma$) confidence, unless otherwise noted.

\section{Observations \& Data Reduction} \label{sec:obs}

\begin{figure*}[t!]
    \centering
    \includegraphics[width=\textwidth]{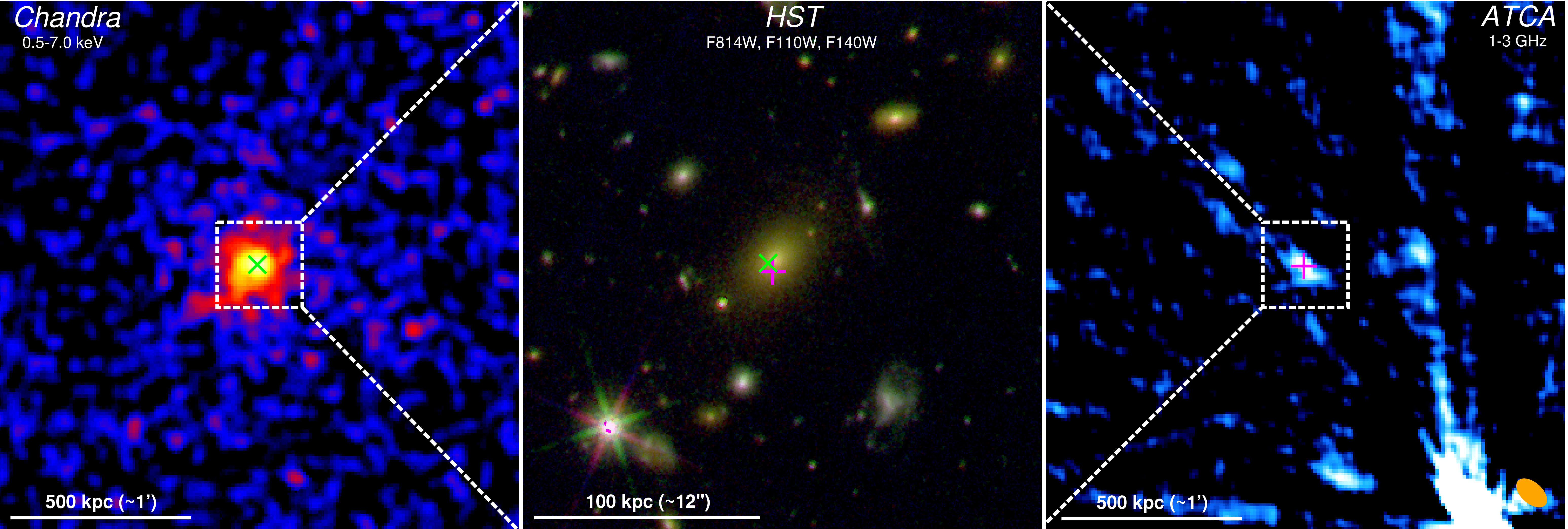}
    \caption{\textit{Left:} Merged \textit{Chandra} X-ray counts image in the broad-band 0.5-7.0~keV. The image is binned such that each pixel is 0\farcs984 on each side and then smoothed with a Gaussian kernel of 4 pixels. The green ``$\times$" shows the location of the X-ray peak, which we use as the center for all X-ray profiles. \textit{Middle:} RGB image of SPT0607 using the HST F140W (red), F110W (green), F814W (blue) filters. The green ``$\times$" shows the location of the X-ray peak and the magenta ``+" shows the location of the radio peak, both of which are coincident with the BCG of SPT0607. \textit{Right:} ATCA 2 GHz radio image with the synthesized beam in orange in the lower right corner. The magenta ``+" shows the location of the radio peak.}
    \label{fig:rgb}
\end{figure*}

In Figure \ref{fig:rgb}, we show the X-ray, optical/infrared (IR), and radio data used in this analysis of SPT0607. On the left and right, we show the \textit{Chandra} X-ray data and ATCA radio data, respectively, and locate the associated peaks in green (X-ray) and magenta (radio). The center panel shows an RGB image using 3 HST filters (F140W, F110W, and F814W), with the same locations of the X-ray and radio peaks overplotted. Both the X-ray and radio peak are coincident with the BCG of SPT0607, as expected for a well-regulated cool core cluster. In the rest of this section, we describe the data and reduction methods used in this paper.

\subsection{Chandra X-ray Observations} \label{subsec:chandra}

SPT0607 was observed with the \textit{Chandra} ACIS-I instrument for a total of 112.5 ksec in January and February 2016. The details of the observations used in this analysis are provided in Table \ref{tab:obs_info}. We reduced and analyzed these data using CIAO \citep[version 4.12;][]{Fruscione2006} and calibration files from CALDB (version 4.9.2.1). All observations were taken in VFAINT mode so we applied additional improved background filtering. We detected and removed point sources using the \texttt{wavdetect} tool and sigma-clipped the light curve at $3\sigma$ with the \texttt{lc\_clean} tool to remove any periods of background flaring from our good time intervals (GTIs). 


At $z = 1.401$ \citep{Khullar2019}, the angular extent of the cluster is relatively small compared to the ACIS-I array, taking up only a single detector chip. Thus, we used an off-source region on the remaining other 3 detectors to produce the background spectra. We extracted source and background X-ray spectra in the 0.5-7.0~keV energy range and used XSPEC (version 12.11.1) for spectral fitting. Spectra were grouped to a minimum of 1 count per bin and $C$-statistic minimization was used for fitting \citep{Cash1979}. We used the XSPEC model \texttt{phabs(apec)}, where the \texttt{phabs} component accounts for absorption in the Milky Way and the \texttt{apec} model accounts for the emission from the intracluster medium. Abundances were taken from \cite{Anders1989}. The absorption column density for the \texttt{phabs} model was free to vary between galactic $N_{HI}$ value, $N_H = 6.78 \times 10^{20}$ cm$^{-2}$ \citep{HI4PI2016}, and the galactic $N_{H,\,\mathrm{tot}}$ value, $N_{H,\,\mathrm{tot}} = N_{HI} + N_{H_2} = 8.33 \times 10^{20}$ cm$^{-2}$ \citep{Willingale2013}. For the cluster emission, we fixed the redshift to $z = 1.401$ and the metallicity to $Z = 0.3 Z_\odot$ given the limited data quality.

\begin{deluxetable}{c c c}

	\caption{\textit{Chandra} Observation Information} \label{tab:obs_info}
	
    \tablehead{\colhead{ObsID} & \colhead{Date} & \colhead{Cleaned Exposure Time} \\ 
    \colhead{} & \colhead{} & \colhead{(ksec)}} 

	\startdata
	17210 & 2016-02-04 & 37.4 \\
	17499 & 2016-01-30 & 39.3 \\
	17500 & 2016-02-20 & 17.8 \\
	18770 & 2016-02-22 & 18.0 \\
    \enddata
    
\end{deluxetable}

\subsection{Optical and Infrared Photometry} \label{subsec:optical_phot}

SPT0607 was observed with the \textit{Hubble Space Telescope} (HST) in four different broad-band filters with Proposal IDs 14252 (PI: V. Strazzullo) and 14677 (PI: T. Schrabback). The cluster was observed in the optical to near-infrared (rest-frame) with the F606W and F814W filters using the Advanced Camera for Surveys (ACS) and with the F110W and F140W filters using the Wide Field Camera 3 (WFC3). The data were reduced using the AstroDrizzle package to remove cosmic rays, perform standard data reduction, and combine images. We utilize the HST photometry primarily to understand the optical spectral energy distribution (SED) of the BCG and calibrate our ground-based spectroscopy. The BCG of SPT0607 is undetected in the bluest filter, F606W, leading to a $1\sigma$ upper limit on the flux of $F_{\lambda, \, \mathrm{F606W}} > 9.1 \times 10^{-20}$~erg~s$^{-1}$~cm$^{-2}$~\AA$^{-1}$.

\subsection{Optical Spectroscopy} \label{subsec:optical_spec}

Optical spectra of potential cluster members of SPT0607 were obtained using the Low Dispersion Survey Spectrograph (LDSS-3C) on the 6.5m Magellan Clay Telescope \citep{Khullar2019}. The VPH-Red grism was used, providing nominal wavelength coverage from 6,000 – 10,000 \AA. With SPT0607 at a redshift of $z = 1.401$, this wavelength coverage provides access to the [\textsc{O~ii}] emission line, which was used to estimate the amount of star formation in the BCG. However, these spectra, initially designed for cluster confirmation by measuring the redshift of potential cluster members, were only wavelength-calibrated and not flux-calibrated. Therefore, in order to obtain a line flux for [\textsc{O~ii}] to estimate star formation rates, we utilized the HST photometry to roughly calibrate the spectrum flux. We first measured an equivalent width from the uncalibrated LDSS-3C spectrum, and then fit the three-band HST photometry to a SED with an old and young stellar population (10 Gyr and 10 Myr, respectively) derived from the \textsc{Starburst99} models \citep{Leitherer1999}. As the BCG in SPT0607 was undetected in the F606W filter, we used only the F814W, F110W, and F140W photometry measurements from HST to fit the SED, which was constrained to within roughly 10\% at the 1$\sigma$ level around the rest-frame wavelength of [\textsc{O~ii}] (see Figure \ref{fig:OII} and Section \ref{subsec:sfr}). This provided a measure of the expected continuum flux at the wavelength of [\textsc{O~ii}], which thus allowed us to convert the equivalent width of the [\textsc{O~ii}] emission line in the LDSS-3C spectrum to a line flux.

\subsection{Radio Observations} \label{subsec:radio_obs}

SPT0607 was observed with the Australia Telescope Compact Array (ATCA) in the 6A configuration in the 1--3~GHz band on 20th August 2016 in seven 20 min visits spread evenly over an 8.5 hour period. These data provide a beam of $6''\times3\farcs5$ at 2~GHz. The data were reduced with the 05/21/2015 release of the {\tt Miriad} software package \citep{Sault1995}. The phase calibrator 0647-475 was used to create the radio maps, with some multi-faceting, but no self-calibration was necessary. The rms value for the resulting image is 23 $\mu$Jy with a dynamic range of $\sim$3000, ensuring sensitivity to extended emission.


\section{Results} \label{sec:results}

\subsection{ICM Properties \& Thermodynamic Profiles} \label{subsec:icm}

In this section, we present the results of the X-ray data analysis whereby we measure the properties of the ICM in SPT0607. We are focused on the core properties of SPT0607, where the impact of AGN feedback is most prevalent, and hence, we measured our radial profiles with respect to the X-ray peak location, as marked in the left and middle panels of Figure \ref{fig:rgb}. As has been noted previously \citep[e.g.][]{McDonald2013,Sanders2018,Ruppin2021}, using a center based on the large scale X-ray centroid, as was done in \citet{McDonald2017} and \citet{Ghirardini2021}, gives a slightly different profile and leads to lower central density and higher central entropy. Additionally, we note that given the relatively high number of counts from SPT0607 ($\sim 700$), our peak location is robust to variations due to noise \citep[e.g.][]{Ruppin2021}.

Due to the high redshift of the source, we make a few conservative assumptions with respect to the temperature profile of the cluster. We first assume that the temperature profile is isothermal, where the temperature is a core-excised temperature measured within a radius $(0.15-1) R_{500}$, using $R_{500} = 0.56$ Mpc from \cite{McDonald2017}. Although this is likely a poor assumption for the true nature of the temperature profile in SPT0607, it provides a strong upper bound on many of our measured thermodynamic properties. In reality, we believe that the cluster has a strong cool core due to the excess surface brightness, radio jet, and lack of significant star formation features in the BCG. We then show in the remainder of this section that we can still recover the features of a strong cool core even with this assumption of an isothermal temperature profile, providing compelling evidence for the cool core nature of this system. After showing that SPT0607 does indeed host a cool core, we also assume a standard cool core temperature profile \citep{Vikhlinin2006}, scaled to the global, core-excised temperature, to obtain a better estimate of the central thermodynamic properties.

\subsubsection{Global Temperature Measurement} \label{subsubsec:temp}

As detailed in Section \ref{subsec:chandra}, we fit the cluster X-ray spectrum in the core-excised region with the simple model \texttt{phabs(apec)} for cluster emission, with the redshift fixed at $z = 1.401$. Cluster metallicity is typically constrained by the highly ionized Fe K-shell lines in X-ray spectra of the ICM, but is poorly constrained in our fits given the high redshift of SPT0607. Thus, we fixed the metallicity at $Z = 0.3 Z_\odot$, motivated by detailed low redshift studies, which find that the average cluster metallicity is roughly a third of the solar value \citep[e.g.][]{Mushotzky1997,DeGrandi2001,Urban2017}, and recent metallicity evolution studies, which show little evolution in the cluster metallicity out to $z \sim 1$ \citep[e.g.][]{McDonald2016b,Flores2021}. The ICM metallicity has been shown to have a weak dependence on temperature \citep[e.g.][]{Fukazawa1998}, and hence, this choice likely has little impact on our measured global temperature. Following the methodology outlined in Section \ref{subsec:chandra}, we find a core-excised temperature of $\langle kT \rangle = 6.75_{-1.51}^{+2.14}$~keV. Using the higher redshift value for SPT0607 of $z = 1.48$ for the cluster redshift (see Section \ref{sec:intro}), we measure a slightly higher core-excised temperature of $\langle kT \rangle = 8.07_{-2.76}^{+6.30}$~keV, but this is consistent with our initial estimate within 1$\sigma$ uncertainty. Using both \textit{Chandra} and \textit{XMM-Newton} data, \citet{Ghirardini2021} found a temperature of $T_0 = 6.0 \pm 0.8$~keV when fitting a Vikhlinin cool core temperature profile, which is consistent with our measurement when considering the differences in the temperature estimates \citep{Vikhlinin2006}.

\subsubsection{Emission Measure and Density Profiles} \label{subsubsec:EM_density}

\begin{figure*}[t!]
    \centering
    \includegraphics[width=17cm]{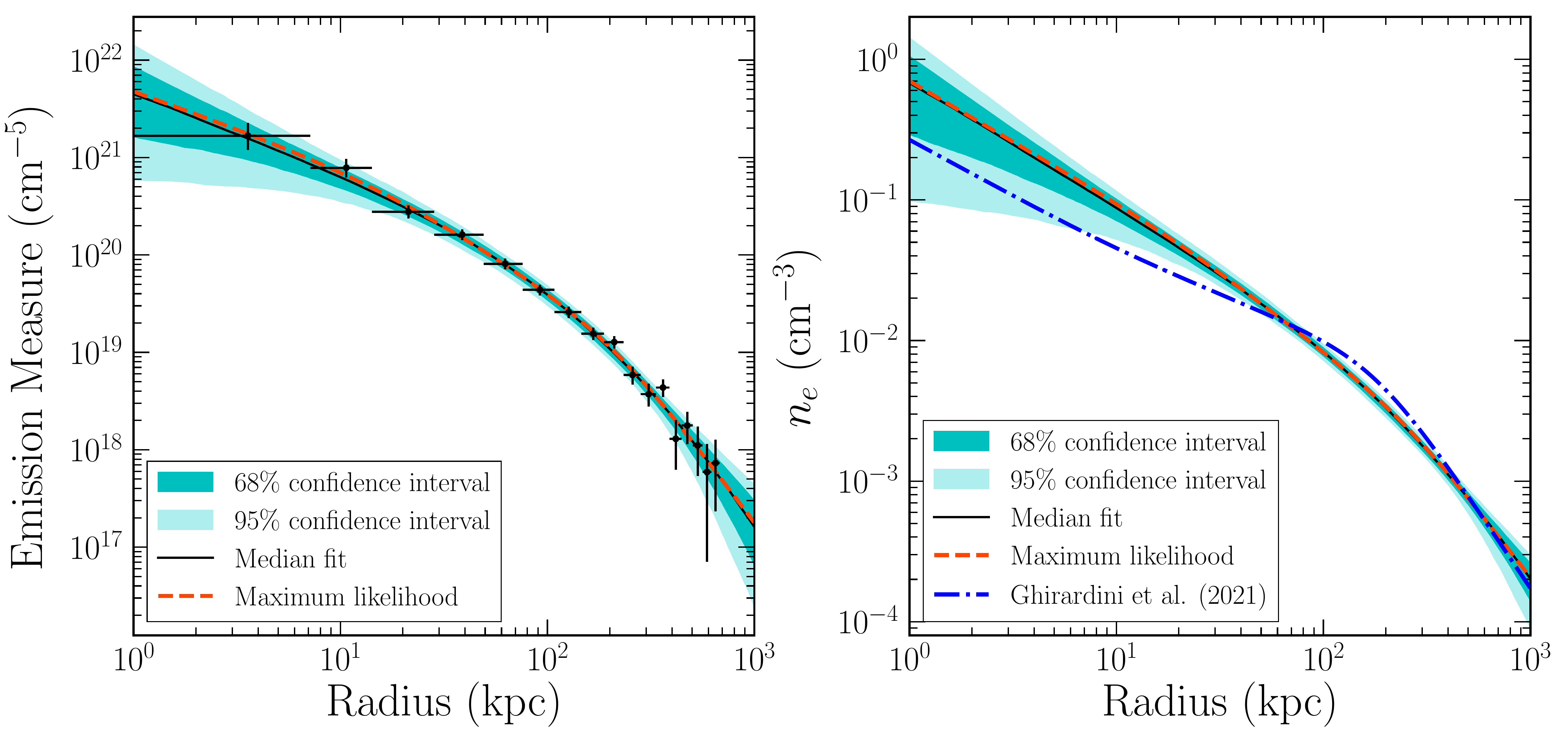}
    \caption{\textit{Left:} The emission measure fit for SPT0607. The emission measure is computed by using the APEC normalization in each of the imaging bins and fitting a projected density profile by integrating along the line of sight through the cluster. The red dashed line shows the maximum likelihood fit, using the Gaussian likelihood given in Equation \ref{eqn:likelihood}. The profile with median fit parameters from the MCMC fit is shown in black, and the confidence interval from the MCMC chain at each radius for 68\% and 95\% confidence is shown in the shaded regions. \textit{Right:} The density profile for SPT0607, computed from the emission measure fit. The maximum likelihood profile is again shown in red, the median MCMC profile is shown in black, and the 68\% and 95\% confidence intervals are shown in the shaded regions. The comparison to the density profile from \cite{Ghirardini2021} is shown in blue. The discrepancy between the two profiles in the core is likely due to our different choice of center (see Section \ref{subsubsec:EM_density})}
    \label{fig:EM_dens}
\end{figure*}

To derive an emission measure from the X-ray data, we extracted a spectrum from each observation in radial bins. We used extraction bins with outer radii defined by  
\begin{equation}
    r_{\mathrm{out}, i} = (a + bi + ci^2 + di^3) R_{500} 
\end{equation}
where the constants $a$, $b$, $c$, and $d$ are as defined in \citet{McDonald2017}, $R_{500} = 560$ kpc \citep{McDonald2017}, and $i = 1, 2, ..., 17$. We use fewer radial annuli than in \citet{McDonald2017} due to poor signal-to-noise in the cluster outskirts for SPT0607. In each radial bin, we fit the spectrum for all 4 observations simultaneously, with all parameters tied across all observations. To derive an emission measure, we simply fix the temperature to the global, core-excised temperature previously described and fit only to the normalization of the \texttt{apec} model. The normalization of the \texttt{apec} model has astrophysical meaning and is given by
\begin{equation}
    \mathrm{norm} = \frac{10^{-14}}{4\pi \left[D_A \left(1+z\right)\right]^2} \int n_e n_H dV,
\end{equation}
where $D_A$ is the angular distance to the source in units of cm, $n_e$ is the electron density in cm$^{-3}$, and $n_H$ is the H density in cm$^{-3}$. Then, by assuming a spherical geometry, the normalization can be related to the emission measure, which is given by 
\begin{equation}
    EM = \int n_e n_H dl,
\end{equation}
where the integral here is along the line of the sight through the cluster. Thus, we can use the \texttt{apec} normalization to obtain the emission measure for each radial bin. Because the normalization measurement is dependent on the temperature we use, we also account for the uncertainty in the temperature measurement by including an additional 10\% uncertainty on each \texttt{apec} normalization measurement (the average difference between the normalization at $\langle kT \rangle$ and the normalization at $\langle kT \rangle \pm 1\sigma$ for the isothermal temperature).
        
To fit the emission measure, we use the modified $\beta$-model \citep{Vikhlinin2006}, whereby the density is given by 
\begin{equation} \label{eqn:modified_beta}
    n_e n_H = n_0^2 \frac{(r / r_c)^{-\alpha}}{(1 + r^2 / r_c^2)^{3\beta - \alpha / 2}} \frac{1}{(1 + r^3 / r_s^3)^{\epsilon/3}},
\end{equation}
where $n_0$ is the central density, $r_c$ and $r_s$ are scaling radii for the cluster core and outskirts, and $r$ is the radial coordinate. This model for the density is then projected and integrated numerically along the line of sight to create an emission measure model. We utilize the Markov Chain Monte Carlo (MCMC) implementation \texttt{emcee} from \cite{ForemanMackey2013} to perform the fitting. We use uniform priors on all parameters and a Gaussian likelihood, given by 
\begin{equation} \label{eqn:likelihood}
    \mathcal{L} = -\frac{1}{2} \chi^2 = -\frac{1}{2} \sum_{i=1}^N \left(\frac{EM_\mathrm{measured} - EM_\mathrm{model}}{\sigma_{EM}}\right)^2,
\end{equation}
where $\sigma_{EM}$ are our errors on the emission measure. We first maximize this likelihood function for our data and then use the maximum likelihood parameters with some scatter as our initial position for the walkers in the MCMC chain. We run the chain with 32 walkers, each for $5 \times 10^5$ chain steps after a burn length of $5 \times 10^4$ chain steps (which is significantly longer than the integrated autocorrelation time of the resulting chain). The resulting fit to the emission measure is shown in the left panel of Figure \ref{fig:EM_dens}. 
        
We can easily turn our emission measure fit into a gas density profile for the cluster since we have fit parameters directly related to the density via Equation \ref{eqn:modified_beta}. For an ionized plasma with a metallicity of $0.3 Z_\odot$, $n_e$ and $n_H$ are related via $n_e = Z n_H$, where $Z = 1.199$ is the average nuclear mass. Likewise, the total gas density of the system can be described by $\rho_g = m_p n_e A / Z$, where $m_p$ is the mass of a proton and $A = 1.397$ is the average nuclear charge. Our density profile is shown in the right panel of Figure \ref{fig:EM_dens}, with a comparison to the density profile from \cite{Ghirardini2021}, which utilizes both \textit{Chandra} and \textit{XMM-Newton} data. \citet{Ghirardini2021} use a large-scale centroid to compute their radial profiles, whereas we choose an X-ray peak approach to capture the core properties. We find decent agreement at the majority of the cluster radii, although our profile predicts a larger overdensity in the cluster core. When using a centroid-based approach (i.e. the \citet{Ghirardini2021} center), we find better agreement between the two profiles, suggesting that the discrepancy in Figure \ref{fig:EM_dens} is due to our choice of using the X-ray peak as the cluster center rather than the large-scale centroid.
   
\subsubsection{Entropy Profile} \label{subsubsec:entropy}
 
\begin{figure}[t!]
    \centering
    \includegraphics[width=8.5cm]{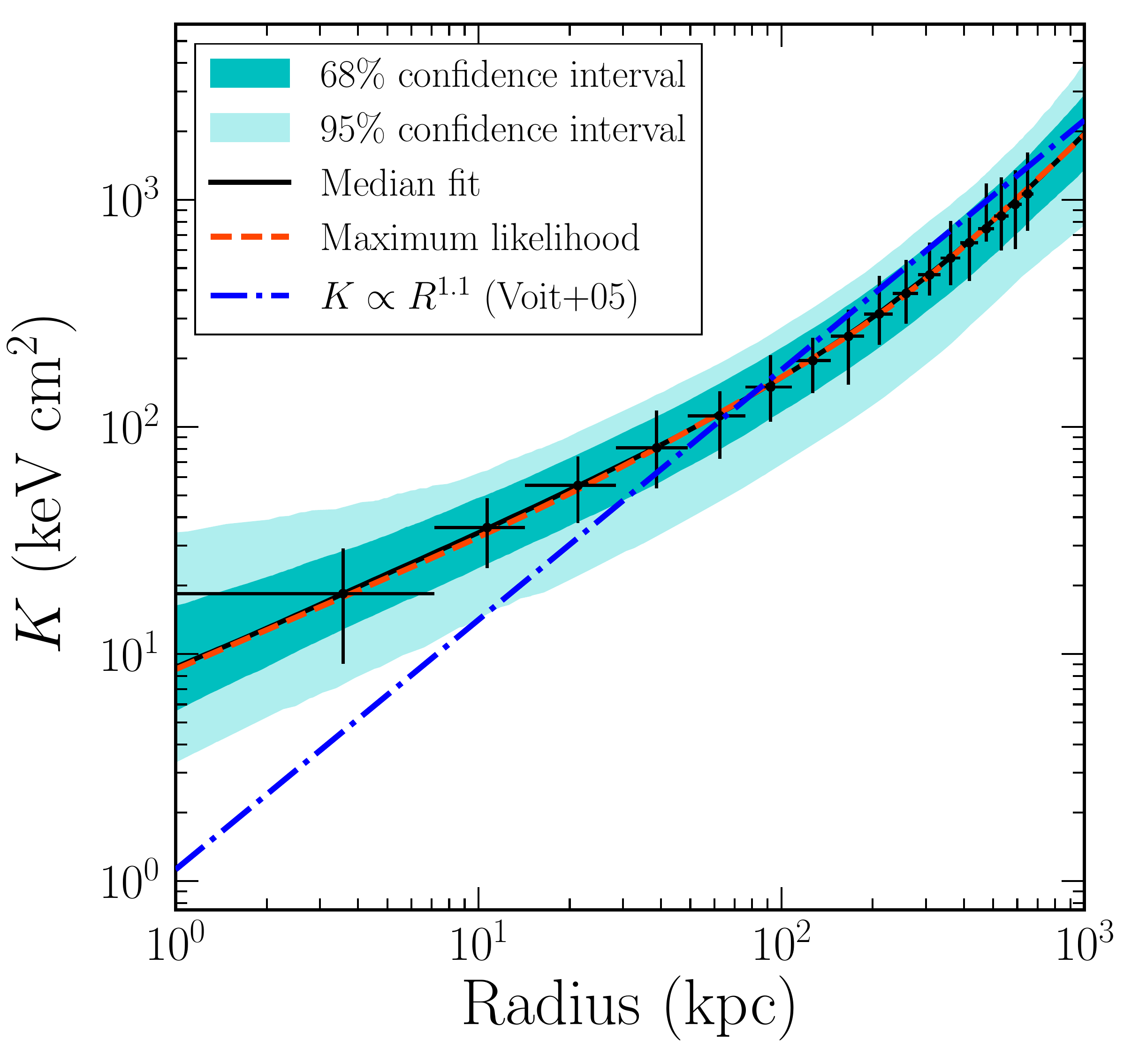}
    \caption{The entropy profile for SPT0607, computed from the derived density profile and an isothermal temperature profile. The analytic profile has been discretized in the same binning scheme used to fit the emission measure data. We find a low entropy core and good agreement in the cluster outskirts with the expected $K \propto R^{1.1}$ relation from \cite{Voit2005}.}
	\label{fig:ent}
\end{figure}

With the density profile for the cluster, we derive an entropy profile, which can both give us insight into the cool core nature of the cluster and trace the thermodynamic history of the ICM \citep{Cavagnolo2009}. Cluster entropy is defined as
\begin{equation}
    K = \frac{kT}{n_e^{2/3}}.
\end{equation}
Assuming an isothermal temperature profile provides an upper limit on the true entropy profile in the core of the cluster. Figure \ref{fig:ent} shows the entropy profile for SPT0607 using the isothermal temperature profile described in Section \ref{subsubsec:temp} and discretizing the entropy in the same bins as we used to measure the emission measure. We find good agreement in the cluster outskirts with the self-similar $K \propto R^{1.1}$ expectation \citep{Voit2005}. In the center, we find slight excess entropy compared to the self-similar expectation, with a central entropy of $K_0 = 18_{-9}^{+11}$~keV~cm$^2$ in the smallest bin ($r \approx 10$ kpc). Thus, even with the most conservative assumption of an isothermal temperature profile, we still recover a low entropy core, consistent with the central entropy in the strong cool cores in the sample from \citet{Hudson2010} ($K_0 \lesssim 22$~keV~cm$^2$). This indicates that SPT0607 is indeed a strong cool core cluster.

To obtain a more accurate estimate of the central entropy, we also computed the entropy profile assuming that the temperature followed the \citet{Vikhlinin2006} cool core profile. Under this assumption, we find a central entropy $K_0 = 10_{-6}^{+3}$~keV~cm$^{-2}$, which is again consistent with a strong cool core in SPT0607.

\subsubsection{Cooling Time} \label{subsubsec:tcool}

\begin{figure}[t!]
    \centering
    \includegraphics[width=8.5cm]{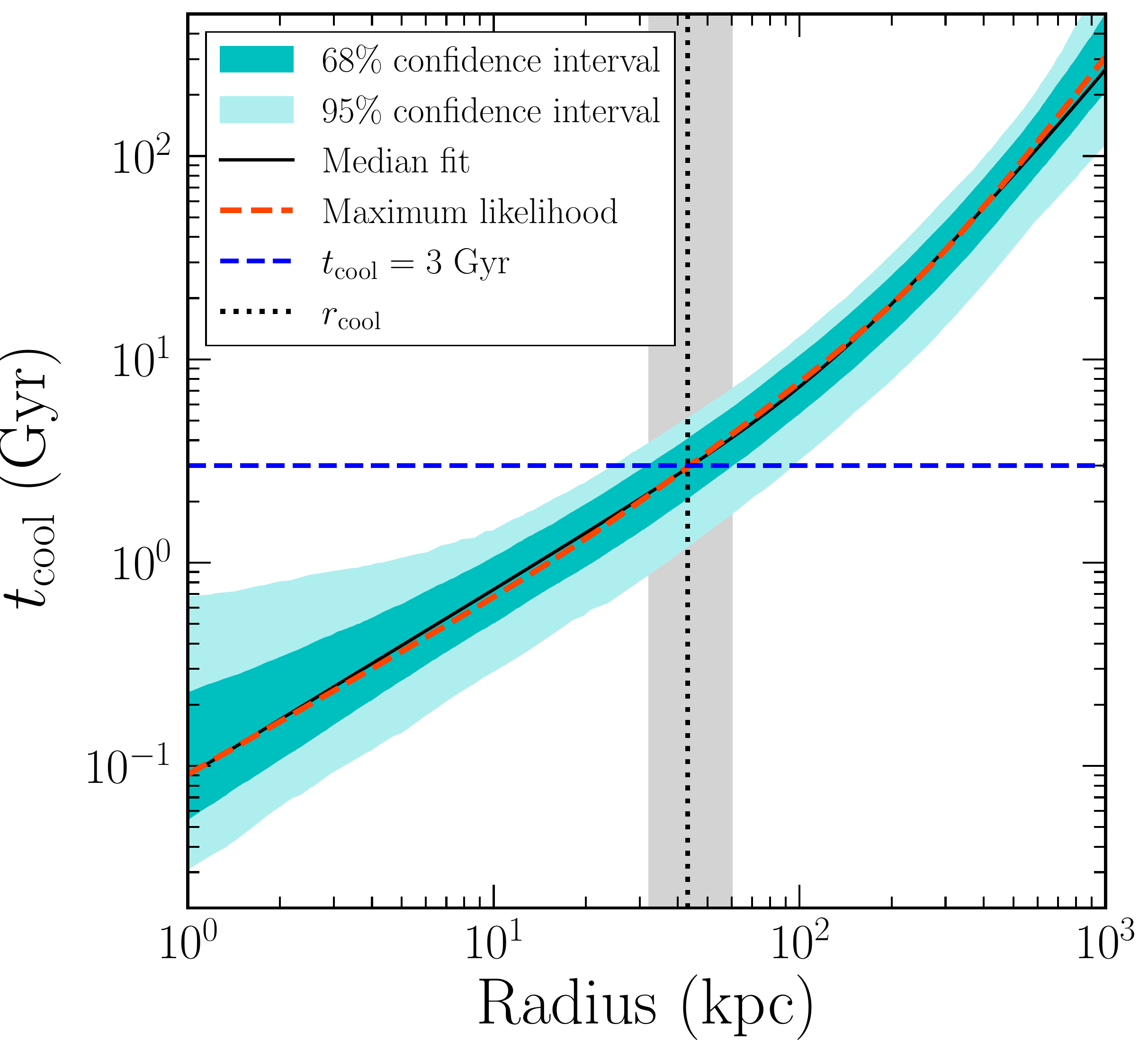}
    \caption{The cooling time profile for SPT0607, computed assuming an isothermal temperature profile and density profiles derived in Section \ref{subsubsec:EM_density}. The radius corresponding to $t_\mathrm{cool} = 3$ Gyr is shown with a blacked dotted line, with the corresponding 68\% confidence interval shown in grey.}
	\label{fig:tcool}
\end{figure}

The last key thermodynamic quantity that we compute is the cooling time, which is used to estimate $r_\mathrm{cool}$ so that we can measure a mass cooling rate to compare with other indicators of cooling to get an idea of the suppression caused by AGN feedback. We compute the cooling time for the cluster using 
\begin{equation} \label{eqn:tcool}
    t_\mathrm{cool} = \frac{3}{2} \frac{(n_e + n_H) kT}{n_e n_H \Lambda(T,Z)},
\end{equation}
where $\Lambda(T,Z)$ is the cooling function for an astrophysical plasma at a temperature $T$ and metallicity $Z$, which we tabulate from \cite{Sutherland1993} for the closest temperature and metallicity for SPT0607. The cooling time profile we derive with an isothermal temperature profile is shown in Figure \ref{fig:tcool}. 

Using this cooling time profile, we measure a cooling radius of $r_\mathrm{cool} = 43_{-11}^{+17}$ kpc, which is defined as the radius at which the cooling time is equal to 3 Gyr. A cooling time of 3 Gyr was chosen as it has been shown to contain the most extended tracers of thermal instabilities in the ICM \citep[e.g.][]{McDonald2010,McDonald2011}. To obtain a mass cooling rate, we then integrate the gas density profile to within the cooling radius and compute the mass cooling rate using 
\begin{equation}
    \dot{M}_\mathrm{cool} = \frac{M_\mathrm{gas}(r < r_\mathrm{cool})}{3 \, \mathrm{Gyr}}.
\end{equation}
From this, we estimate from the X-ray analysis that the expected mass cooling rate is $\dot{M}_\mathrm{cool} = 100_{-60}^{+90} \, M_\odot$ yr$^{-1}$. Similarly to the central entropy, we also compute this value using a scaled version of the universal cool core temperature profile and find consistent mass cooling rates under that assumption.

\subsection{Radio Power} \label{subsec:radio}

We utilize ATCA 2.1 GHz observations of SPT0607 to determine the total radio power associated with the BCG in SPT0607. The jet from the BCG is unresolved, and we measure an integrated flux using CASA \citep{McMullin2007} of $S_{2.1 \, \mathrm{GHz}} = 0.23 \pm 0.11$ mJy within an ovular aperture equal to the beam size centered on the radio peak. This corresponds to a 2.1 GHz radio luminosity of $L_{2.1 \, \mathrm{GHz}} = (2.3 \pm 1.1) \times 10^{24}$ W Hz$^{-1}$. We then estimate the radio power using
\begin{equation}
    P_{\nu_0} = 4 \pi D_L^2 (1 + z)^{\alpha - 1} S_{\nu_0} \nu_0,
\end{equation}
from \cite{Cavagnolo2010}, where $\nu_0$ is the observed frequency (2.1 GHz), $S_{\nu_0}$ is the flux density at the observed frequency, $D_L$ is the luminosity distance, and $\alpha$ is the spectral index. Since we only have data at one frequency from ATCA, we cannot measure the spectral index, but instead adopt a typical value for extragalactic radio galaxies of $\alpha = 0.8$ as in \cite{Cavagnolo2010}. Using a spectral index of $\alpha = 0.8$, we find a radio power of $P_{2.1 \, \mathrm{GHz}} = (4.8 \pm 2.4) \times 10^{40}$~erg~s$^{-1}$. 

To compare the power of the radio jet in the BCG to the amount of cooling expected in the ICM, we use the scaling relation from \cite{Cavagnolo2010} to convert the measured radio power to a jet power. We first use the same spectral index to convert the observed 2.1 GHz power to a 1.4 GHz power, which can then be directly converted to jet power using Equation (1) of \cite{Cavagnolo2010} given by 
\begin{equation}
    \log P_\mathrm{cav} = (0.75 \pm 0.14) \log P_{1.4} + (1.91 \pm 0.18),
\end{equation}
where $P_\mathrm{cav}$ is in units of $10^{42}$~erg~s$^{-1}$ and $P_{1.4}$ is in units of $10^{40}$~erg~s$^{-1}$. We find a jet power of $P_\mathrm{cav} = 3.2_{-1.3}^{+2.1} \times 10^{44}$~erg~s$^{-1}$ using this scaling relation. To compare the heating from the radio jet to the cooling of the ICM, we compute the X-ray cooling luminosity of the ICM within $r_\mathrm{cool}$, using our derived value of $r_\mathrm{cool}$ from Section \ref{subsubsec:tcool}. We find an unabsorbed X-ray cooling luminosity of $L_\mathrm{cool} = 1.9_{-0.5}^{+0.2} \times 10^{44}$~erg~s$^{-1}$ in the 0.01-100~keV band, which is identical to the radio jet power within 1$\sigma$ confidence. This is consistent with the radio BCG power versus X-ray cooling luminosity found in a large sample of low redshift clusters in \citet{Hogan2015}, as well as with the lack of a significant redshift evolution in $P_\mathrm{cav} / L_\mathrm{cool}$ for clusters out to $z \sim 1.3$ in \citet{Ruppin2022}. The implications of these findings on the regulation of cooling in SPT0607 by radio-mode AGN feedback are discussed further in Section \ref{sec:discussion}.

\subsection{Regulated Star Formation in the BCG} \label{subsec:sfr}

Using the LDSS-3C optical spectrum from the Magellan Clay telescope, we estimate the star formation rate (SFR) in the BCG by measuring a luminosity of the [O\,\textsc{ii}] $\lambda\lambda 3727,3729$ \AA\, doublet. The [O\,\textsc{ii}] emission feature is a useful indicator of star formation \citep[e.g.][]{Kennicutt1998,Kewley2004}, especially in the high-redshift universe because it has a similar ionization energy to hydrogen, but, unlike the $H\alpha$ transition, is not redshifted out of the optical band. The [O\,\textsc{ii}] emission traces warm gas with $T \sim 10^4$ K around young O and B stars, thus tracing instantaneous star formation on timescales on the order of $\sim 10$ Myr. However, SFRs derived from [O\,\textsc{ii}] emission line are more dependent on dust, metallicity, and ionization than other tracers like H$\alpha$, UV, and far-IR luminosities \citep[e.g.][]{Rosa-Gonzalez2002,Kewley2004,Moustakas2006}, which we cannot accurately determine with current data on SPT0607. AGN can also excite [O\,\textsc{ii}] in the nuclei of galaxies, but the AGN in SPT0607 is radiatively inefficient and weak in X-ray emission. Thus, we do not expect the central AGN to be contributing significantly to the [O\,\textsc{ii}] emission in SPT0607 and can safely attribute the majority of the [O\,\textsc{ii}] emission to star formation.
    
\begin{figure}[t!]
    \centering
    \includegraphics[width=8.6cm]{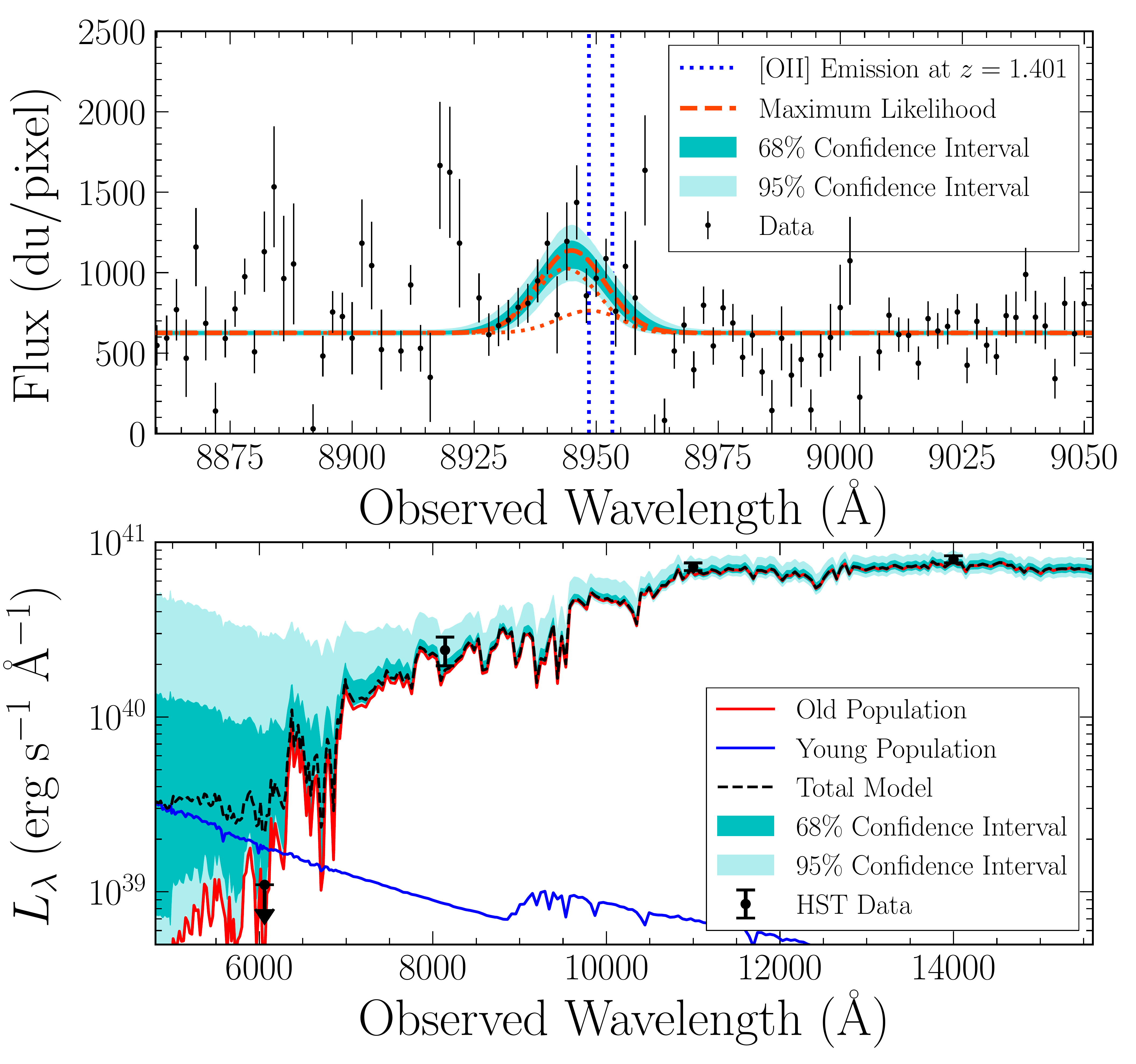}
    \caption{\textit{Top}: Fit to the wavelength-calibrated LDSS-3C spectrum around the [\textsc{O~ii}] emission feature. The maximum likelihood fit is shown as a red dashed line, with the two individual Gaussian components shown with red dotted lines. Confidence intervals are shown in green. The observed wavelength of the [\textsc{O~ii}] doublet is shown with blue dotted lines. We allow for some systematic offset from the observed wavelength to account for motion within the cluster. \textit{Bottom}: Fit to the three-band HST photometry using a simple young and old stellar population model from \textsc{Starburst99} \citep{Leitherer1999}. The total model is shown in black, with confidence intervals in green. The young and old stellar population contributions are shown in blue and red, respectively. A 1$\sigma$ upper bound from the F606W filter is also shown, although this is not used in the fitting procedure. The SED fit is used to obtain a continuum flux at the wavelength of [\textsc{O~ii}], with which we can combine the equivalent width measurement from the top panel to determine the [\textsc{O~ii}] line flux.}
	\label{fig:OII}
\end{figure}

We fit the LDSS-3C spectrum within 100 \AA\, on either side of the expected [\textsc{O~ii}] emission feature with a constant to estimate the continuum and doublet Gaussian feature for the [\textsc{O~ii}] line. We fix the redshift at $z = 1.401$ for the cluster, and allowed the line centers to vary within 500 km s$^{-1}$ of the atomic value to account for peculiar motions in the cluster. We restrict the width of the line to be less than 500 km s$^{-1}$ to account for turbulent motions broadening the line. We tie the widths of the two Gaussian components together and allowed their line ratio to be free. We use the \texttt{emcee} package \citep{ForemanMackey2013} with a Gaussian likelihood and uniform, uninformative priors to fit the spectrum using an MCMC approach with 32 walkers, 50,000 chain steps per walker, and a burn length of 5,000 chain steps per walker (which is significantly longer than the integrated autocorrelation time of the resulting chain). The result of the fit is shown in the top panel of Figure \ref{fig:OII}. We detect a relatively weak emission feature in [\textsc{O~ii}] with a velocity offset of $v = -200 \pm 60$~km~s$^{-1}$, a line width of $230 \pm 40$~km~s$^{-1}$, and a rest-frame equivalent width (EW) of $EW_\mathrm{[OII]} = 6.0 \pm 0.9$~\AA. This equivalent width is then turned into a line flux using the flux-calibrated HST photometry to model the continuum SED, as shown in the bottom panel of Figure \ref{fig:OII} and detailed in Section \ref{subsec:optical_spec}.

From this calibration, we measure an [\textsc{O~ii}] luminosity of $L_\mathrm{[OII]} = 1.3_{-0.2}^{+0.3} \times 10^{41}$~erg~s$^{-1}$, which has not been corrected for extinction. We account for extinction by folding in uncertainty on $E(B-V)$ by assuming a uniform distribution between $E(B-V) = 0$ (i.e. dust-free) and $E(B-V) = 0.3$. Using Equations (10) and (17) of \cite{Kewley2004}, we convert our observed [\textsc{O~ii}] luminosity to a SFR (assuming a solar value of $\log(O/H) + 12 = 8.9$). From our MCMC chains from fitting the line and folding in the uniform distribution of $E(B-V)$, we obtain an extinction-corrected star formation rate of SFR$_\textsc{[O\,ii]} = 1.7_{-0.6}^{+1.0} \, M_\odot$~yr$^{-1}$. This value is more than two orders of magnitude lower than the cooling rate we measure in the X-ray band, indicating that the cooling in SPT0607 is well-regulated by AGN feedback. Likewise, this star formation rate is comparable to low-redshift samples of BCGs with little on-going star formation as measured with H$\alpha$ and other SFR indicators \citep[e.g.][]{Crawford1999,McDonald2010}. This thus adds to the evidence that SPT0607 is a high-redshift analog of the large population of relaxed, low-redshift clusters with well-regulated star formation and ICM cooling by AGN feedback.


\section{Discussion} \label{sec:discussion}

From the analysis of X-ray, optical, and radio observations, SPT0607 clearly hosts a strong cool core with AGN feedback offsetting the cooling from the ICM, as is common place in low redshift galaxy clusters. An overview of properties of the cluster and BCG derived in this work are given in Table \ref{tab:props}, highlighting the low central entropy, similarity of the radio cavity power and cooling luminosity, and the SFR that is $\sim$1\% of the predicted mass cooling rate. In the remainder of this section, we discuss the implications that these findings have on our understanding of high redshift clusters and the evolution of AGN feedback.

\begin{deluxetable}{c c}

	\caption{Summary of Cluster and BCG Properties} \label{tab:props}
	
    \tablehead{\colhead{BCG Property} & \colhead{Value}} 

	\startdata
	Central Entropy & $K_0 = 18_{-9}^{+11}$~keV~cm$^2$ \\
	X-ray Mass Cooling Rate & $\dot{M}_\mathrm{cool} = 100_{-60}^{+90} \, M_\odot$~yr$^{-1}$ \\
	X-ray Cooling Luminosity & $L_\mathrm{cool} = 1.9_{-0.5}^{+0.2} \times 10^{44}$~erg~s$^{-1}$ \\
	Radio Jet Power & $P_\mathrm{cav} = 3.2_{-1.3}^{+2.1} \times 10^{44}$~erg~s$^{-1}$ \\
	Star Formation Rate & $1.7_{-0.6}^{+1.0} \, M_\odot$~yr$^{-1}$ \\
    \enddata
    
\end{deluxetable}

\subsection{Constraints on the Onset of Radio-Mode Feedback}

At low redshifts, radio-mode AGN feedback, whereby the central AGN accretes mass at a low rate and launches radio jets that deposit large amounts of mechanical energy into the ICM, is the main mechanism by which runaway ICM cooling is prevented in cool core clusters \citep[e.g.][]{Birzan2004,Dunn2006,Rafferty2006}. Through multi-wavelength observations, we have shown that SPT0607 has well-regulated radio-mode feedback from its BCG and, to our knowledge, is the highest redshift cluster with these properties known to date. As such, it provides one of the strongest constraints to date on the onset of AGN feedback in galaxy clusters. 

Simulations and theoretical models of the evolution of AGN feedback and supermassive black hole growth suggest that on average AGN in cluster environments should transition from quasar-mode feedback at early times, where the black hole is accreting at higher rates and the accretion process is radiatively efficient, to radio-mode feedback at late times \citep[e.g.][]{Churazov2005,Croton2006}. Recent simulations suggest that this transition should take on the order of 1-2 Gyr to occur in BCGs in cool core clusters \citep[e.g.][]{Qiu2019}. Indeed, at low redshifts, only on the order of 1-2\% of clusters are observed to have a X-ray bright central AGN, which is expected for radiatively efficient accretion in the BCG and quasar-mode feedback \citep[e.g.][]{Green2017,Somboonpanyakul2021}. SPT0607 has well-regulated radio-mode feedback from its BCG, suggesting that the radio-mode feedback must be present and a dominant form of AGN feedback in some clusters out to at least $z = 1.4$. Whether this is the dominant mechanism of feedback in most high redshift systems is a question that still remains to be answered with a more complete sample of radio and X-ray observations of high redshift clusters. However, we can use SPT0607 to place constraints on the minimum redshift at which AGN feedback must have turned on in clusters; under the assumption that BCGs are dominated by radiatively efficient accretion during the first 1-2 Gyrs \citep{Qiu2019}, the lowest redshifts at which the AGN feedback process could have began in SPT0607 is $z \sim 1.9-2.6$.

Previously, studies of X-ray cavities from jet-powered bubbles in the ICM have shown there is little evolution in the properties of radio-mode feedback from the local universe back to $z \sim 0.8$ \citep{Hlavacek-Larrondo2012,Hlavacek-Larrondo2015}. Additionally, the discovery of more distant cool core clusters with central radio sources capable of balancing ICM cooling, such as WARPJ1415.1+3612, have extended these findings out to $z \sim 1$ \citep{Santos2012}. With SPT0607, we can extend this relation even further out to $z = 1.4$. However, it is still unclear when radio-mode feedback was established in galaxy clusters and how the fraction of clusters with well-regulated AGN feedback has evolved out to high redshifts. The next generation X-ray observatories will target this question by probing the ICM in the most distant clusters, with the ability to detect cluster emission out to $z \sim 2-3$ \citep{Barret2020}. With many more systems, we will be able to get a better handle on the evolution of radio-mode feedback and the AGN duty cycle in high redshift clusters. For now, at $z = 1.401$, SPT0607 provides the furthest constraint on the onset of radio-mode feedback in cool core clusters. 

\subsection{Star Formation in BCGs at High Redshift}

Star formation in the BCGs in cool core clusters is a critical piece of the AGN feedback process as it acts as a probe of the balance between heating by AGN feedback and cooling in the ICM. Various works have found that both the star formation rate and specific star formation rate of BCGs increase as a function of increasing redshift \citep[e.g.][]{Webb2015,McDonald2016a,Bonaventura2017}. However, the nature of star forming BCGs seems to have changed with redshift. In particular, \citet{McDonald2016a} found that there was a transition in the fuel supply of the BCG, namely that high-redshift clusters out to $z \sim 1.2$ with highly star forming BCGs were almost always disturbed clusters. This suggests that gas-rich mergers are responsible for runaway cooling and star formation in high-redshift systems, rather than cooling flows from a lack of heating from AGN feedback, as was recently observed in the $z \sim 1.7$ system SpARCS1049 \citep{Hlavacek-Larrondo2020}. However, at low redshifts, star forming BCGs are predominantly found in relaxed systems, indicating that star formation in BCGs at low redshifts is commonly driven by cooling of the ICM and regulated by AGN feedback. With multi-wavelength observations of SPT0607, we have found that this high-redshift, relaxed cluster hosts a BCG with very little star formation. The BCG also shows no noticeable morphological features in the 3-band HST images that suggest any recent mergers of interactions. These findings thus agree with the idea of a transitioning fuel supply for BCG star formation at high redshift, where the majority of the fuel for star formation in high-redshift systems comes from gas rich mergers as clusters are assembling. SPT0607 supports this picture out to $z \sim 1.4$ and suggests that the early onset of AGN feedback provides sufficient heating to offset direct cooling from the ICM into stars at high redshift.


\section{Summary} \label{sec:summary}

We have presented a multi-wavelength analysis of one of the most distant SPT-selected clusters, SPT0607 at a redshift of $z = 1.401$. Through analysis of \textit{Chandra} X-ray data, we found that SPT0607 has a strong cool core, as evidenced by both an increase in central gas density and a low entropy core as measured from the X-ray peak. These results follow from our conservative assumption of an isothermal temperature profile; in reality, we expect the central temperature of SPT0607 to drop in the center, which gives an even lower entropy core when assumed. 

As shown in Figure \ref{fig:rgb}, the core of SPT0607 is coincident with the BCG, which harbors a radio jet detected with ATCA at 2.1 GHz. Despite having a dense and cool core, we measure a star formation rate in the BCG of SPT0607 of SFR$_\textsc{[O\,ii]} = 1.7_{-0.6}^{+1.0}$ $M_\odot$~yr$^{-1}$ using measurements of the [\textsc{O~ii}] emission line from optical spectroscopy with the LDSS-3C instrument on the 6.5m Magellan Clay telescope. This star formation rate is roughly 1\% of the expected mass cooling rate of the ICM of $\dot{M}_\mathrm{cool} = 100_{-60}^{+90}$ $M_\odot$~yr$^{-1}$ from our X-ray measurements. Similarly, we measure a cavity power from the radio jet of $P_\mathrm{cav} = 3.2_{-1.3}^{+2.1} \times 10^{44}$~erg~s$^{-1}$, which is consistent with the X-ray cooling luminosity. This indicates that the BCG in SPT0607 is providing radio-mode feedback to offset the cooling from the ICM. This phenomenon is commonplace at low redshift, but as one of the most distant clusters known to date, the regulation of cooling and AGN feedback in SPT0607 gives the strongest constraints on the onset of radio-mode AGN feedback in galaxy clusters to date. 


\medskip
\noindent
The South Pole Telescope program is supported by the National Science Foundation (NSF) through grants PLR-1248097 and OPP-1852617. Partial support is also provided by the NSF Physics Frontier Center grant PHY-1125897 to the Kavli Institute of Cosmological Physics at the University of Chicago, the Kavli Foundation, and the Gordon and Betty Moore Foundation through grant GBMF\#947 to the University of Chicago. Argonne National Laboratory’s work was supported by the U.S. Department of Energy, Office of Science, Office of High Energy Physics, under contract DE-AC02-06CH11357. The Melbourne group acknowledges support from the Australian Research Council’s Discovery Projects scheme (DP200101068).

All of the HST data used in this paper can be found in MAST: \dataset[10.17909/e40m-z102]{http://dx.doi.org/10.17909/e40m-z102}.

\facilities{CXO, HST, Magellan, ATCA, NSF/US Department of Energy 10m South Pole Telescope (SPT-SZ)}

\software{CIAO \citep{Fruscione2006},
XSPEC \citep{Arnaud1996}, 
CASA \citep{McMullin2007},
STARBURST99 \citep{Leitherer1999},
Astropy \citep{Astropy2013, Astropy2018},
Matplotlib \citep{Hunter2007},
NumPy \citep{vanderWalt2011}}


\bibliography{SPT0607_biblio.bib}

\end{document}